\documentclass[prd,showpacs,showkeys,preprint]{revtex4}

\usepackage{amsmath}
\usepackage{amssymb}
\usepackage{yfonts}[1988/10/03]
\usepackage{graphicx}
\newcommand {\beq}{\begin{equation}}
\newcommand {\eeq}{\end{equation}}
\newcommand {\beqa}{\begin{eqnarray}}
\newcommand {\eeqa}{\end{eqnarray}}

\begin{document}

\title{ On the Lorentz invariance of the Square root Klein-Gordon (Salpeter)Equation }
\author{  M. J. Kazemi $^{1}$\footnote{kazemi.j.m@gmail.com}, M. H. Barati$^{2}$\footnote{mohbarati14@gmail.com}, Jafar Khodagholizadeh $^{3}$\footnote{Lecturer in Farhangian University, Tehran,Iran. Email: gholizadeh@ipm.ir}and Alireza
 Babazadeh$^{4}$\footnote{arbabazadeh@yahoo.com}}
\affiliation{$^{1}$ Department of Physics, Shahid Beheshti University, G. C., Evin,Tehran 19839, Iran.\\
$^{2}$
Department of Physics, kharazmi University, Tehran, Iran.\\
$^{3}$School of physics, Institute for research in fundamental sciences (IPM), Tehran, Iran\\
$^{4}$Optics, Laser and Photonics Insttitute , Amirkabir University of Technology, Valiasr Ave.,Tehran, Iran.
}

\begin{abstract}
We show that the Born's rule is incompatible with Lorentz symmetry of the Square Root Klein-Gordon equation (Salpeter equation). It has been demonstrated that the Born rule must be modified in relativistic regime if one wishes to keep the Salpeter equation  as the correct equation for describing quantum behavior.
\end{abstract}

\pacs{ }
%\preprint{***}
\maketitle
\section{Introduction}
The simplest relativistic generalization of the Schrodinger equation  could be derived from direct substitution of $p\rightarrow i\hbar \nabla $ and $ E\rightarrow i\hbar \frac{\partial}{\partial t}$  into the relativistic energy-momentum relation  $  E=\sqrt{c^2 p^2+m_0^2 c^4}$. Such procedure leads to
\begin{equation}\label{1}
i h \frac{\partial \psi}{\partial t}=mc^{2}\sum_{k=0}^{\infty}(\dfrac{i\hbar}{mc})^{2k}\begin{pmatrix}
  1/2 \\
  k
\end{pmatrix}\nabla^{2k}\psi
\end{equation}
this equation is  so called "Square Root Klein-Gordon Equation"(SRKG equation) or "Salpeter Equation". The integral-differential form of this equation can be used in order to circumvent the divergence of above expansion:
\begin{equation}
i \hbar \frac{\partial \psi(\textbf{x},t)}{\partial t}=\int K(\textbf{x}-\textbf{x}^{'})\psi(\textbf{x}^{'})d\textbf{x}^{'}
\end{equation}
where the integral kernel,$\textit{K}$  can be expressed  as follows:
\begin{equation}
K(\textbf{z})= -\frac{mc^{2}}{2\pi^{2}}\frac{K_{2}(|\textbf{z}|/l_{c})}{|\textbf{z}|/l_{c}}
\end{equation}
Where $K_{\nu}(|\textbf{z}|/l_{c})$  is the modified Bessel function (Macdonald function) and $l_c$  is the Compton wavelength. Historical background of the equation (\ref{1}) dates back to the early years of the relativistic quantum mechanics. In 1927,  Wayl proposed using  square root operator, $\sqrt{-c^2 \hbar^2 \nabla^2+m_0^2 c^4}$ to formulate the relativistic quantum mechanics \cite{1}, however he didn't develop his idea to a comprehensive theory. On the other hand, other pioneers of quantum mechanics used different methods to formulate Relativistic Quantum Mechanics which led to the Dirac and the Klein- Gordon equations. Eventually, the square root Klein- Gordon equation often did not  accepte in the growing formulation of Relativistic Quantum Mechanics. But in recent years, theoretical characteristics and integral representations of this equation have been the matter of interest \cite{2,3,4,5,6,7}. Moreover, this equation recently was used to describe some phenomena and problems in relativistic regime such as: Relativistic Harmonic Oscillator \cite{8,9,10}, waves in Relativistic Quantum Plasma \cite{11}, Relativistic Bound state (quark-antiquark-gluon systems)\cite{12,13,14,15,16,17,18}, and Relativistic Bohmian Mechanics \cite{19}. We particularly point out the consistency between the results of this equation with the experimental spectrum of the mesonic atoms \cite{12}. Furthermore, since this equation is first order with respect to time, we can use the Born rule $ (\rho_B= |\psi|^2 $) to interpret the wave function  and hence the problem of negative probability density of the Klein-Gordon equation would not arise. In this regard, the current density and the Born probability density is obtained as follows:\cite{3}
\begin{eqnarray}\label{4}
\rho_{B}= |\psi|^2
\end{eqnarray}
\begin{eqnarray}\label{5}
 \bold{ \textit{J}} _{B}=-\dfrac{imc^{2}}{\hbar}\sum_{k=1}^{\infty}\dfrac{(2k-3)!!}{(2k)!!}(\dfrac{\hbar}{mc})^{2k}\nonumber\\ \times \sum_{l=0}^{2k-1}(-1)^{l}\bold{\nabla}^{l}\psi^{\star}\bold{\nabla}^{2k-2l-1}\psi
\end{eqnarray}
In which, $ \rho_B $ is the Born probability density and $ \textbf{J}_B $ is the corresponding current density. In fact this  probabilistic interpretation possibility of the wave function has been often a motivation for utilizing the square root Klein-Gordon equation.

On the other hand,the Lorentz invariance of this equation has often been under discussion. Because of its high derivatives, checking Lorentz invariance of this equation is so complicated and cannot be specified easily, . And also , because of the inequality of time and space derivatives, this equation has been mistakenly accepted as  a frame dependent equation and incompatible with special relativity \cite{16},\cite{20,21,22,23,24,25}. It should be noted that for checking the Lorentz invariance of an equation need to be determined Lorentz transformation of all quantity in the equation; and one cannot give opinions about the Lorentz invariance of an equation only based on the Lorentz transformation of the differential operators. On the other hand the Lorentz transformation of every quantity must be determined by their physical definitions. Thus Lorentz transformation of complex quantities such as wave function must be determined based on their physical interpretation and their relations to the observable quantities. For this purpose by accepting the probabilistic interpretation of the wave function, the Lorentz transformation of the wave function must be determined by its relation to position probability density $ \rho $ . Explicitly, Lorentz transformation of the wave function should be considered in a way that probability density ,$ \rho $ and the probability current density, $ \textbf{J }$ are altogether transformed as one four-vector.Therefore, it should be noted that before the wave function is interpreted - which determines its Lorentz transformation - the square root Klein-Gordon is neither Lorentz invariant nor Lorentz non-invariant.

In fact the Lorentz invariance of the square root Klein-Gordon equation was already shown; considering the wave function as scalar (in the absence of interaction). But scalar wave function is inconsistent with Born interpretation because $ \rho_{B}=\vert\psi\vert^{2} $ is the first component of the probability current four-vector and cannot be scalar. Now if we accept the Born rule for the probabilistic interpretation of wave function, then the common proofs for the Lorentz invariance of the square root Klein-Gordon equation are incomplete because they are all under the condition of the wave function being scalar.
The question that arises at this point is whether or not we could find any Lorentz transformation for the wave function that leads to: 1) Lorentz invariance of the square root Klein-Gordon equation; and 2) $ \rho_{B} $ and  $ J_{B} $ are altogether transformed as one four-vector. In the next section we will show that there does not exist such transformation; and because the two above conditions cannot be simultaneously satisfied, the Born rule is completely inconsistent with the Lorentz symmetry of the square root Klein-Gordon equation.
\section{prove of incompatibility of the born rule with lorentz symmetry of the square root Klein-Gordon equation}
In this section we will present a simple counterexample to show there is no transformation for the wave function that leads to the Lorentz invariance of all three equations (\ref{1}),(\ref{4}) and (\ref{5}).
 In this regard we consider the wave function as a superposition of two plane waves in two inertial reference frames $\textit{S}$ and $ \textit{S}^{'} $ (in one dimension):
\begin{eqnarray}
\psi(x,t)=\sum_{i=1}^{2}A_{i}e^{\dfrac{i}{\hbar}(p_{i}x-E_{i}t)}
\end{eqnarray}
\begin{eqnarray}
\psi^{'}(x^{'},t^{'})=\sum_{i=1}^{2}A_{i}^{'}e^{\dfrac{i}{\hbar}(p_{i}^{'}x^{'}-E_{i}^{'}t^{'})}
\end{eqnarray}
Where $ E_i=p_i^0=\sqrt{p_i^2 c^2+m_0^2 c^4} $, It should be noted that this choice for time evaluation of the wave functions ensure the establishment of  square root  Klein-Gordon equation in both frameworks. Applying equations (\ref{4}) ,(\ref{5}) for above wave functions leads to:
\begin{eqnarray}
\rho_{B}=\sum_{i,j=1}^{2}|A_{i}||A_{j}|\cos((p_{i}^{\mu}-p_{j}^{\mu})x_{\mu}+\delta_{ij})
\end{eqnarray}
\begin{eqnarray}
\rho_{B}^{'}=\sum_{i,j=1}^{2}|A_{i}^{'}||A_{j}^{'}|\cos((p_{i}^{\mu}-p_{j}^{\mu})x_{\mu}+\delta_{ij}^{'})
\end{eqnarray}
 \begin{eqnarray}
 J_{B}=\sum_{i,j=1}^{2}|A_{i}||A_{j}|U_{ij}\cos((p_{i}^{\mu}-p_{j}^{\mu})x_{\mu}+\delta_{ij})
 \end{eqnarray}
 \begin{eqnarray}
 J_{B}^{'}=\sum_{i,j=1}^{2}|A_{i}^{'}||A_{j}^{'}|U_{ij}^{'}\cos((p_{i}^{\mu}-p_{j}^{\mu})x_{\mu}+\delta_{ij}^{'})
 \end{eqnarray}
 Where $ \delta_{ij} $  is the phase difference between $ A_{i} $ and $ A_{j} $ and $ U_{ij} $ be defined as follows:
 \begin{eqnarray}\nonumber
 U_{ij}=\dfrac{p_{i}+p_{j}}{E_{i}+E_{j}}c^{2}
 \end{eqnarray}
 and prime quantities are similarly defined in $ S^{'} $. Lorentz transformation of probability density ,$ \rho^{'}=\gamma(\rho_{B}-\dfrac{v}{c^{2}}J_{B}) $ leads to following equations:
 \begin{eqnarray}\label{12}
 |A_{1}^{'}||A_{2}^{'}|=\alpha_{12}|A_{1}||A_{2}|
 \end{eqnarray}
 \begin{eqnarray}\label{13}
  \sum_{i=1}^{2}(|A_{i}^{'}|^{2}-\alpha_{ii}|A_{i}|^{2})=0
\end{eqnarray}
where $  \alpha_{ij}=\gamma(1-(v U_{ij})/c^2 ) $. Also current Lorentz transformation,$  J_{B}^{'}= \gamma (J_B-v \rho_B ) $, leads to:
 \begin{eqnarray}\label{14}
  |A_{1}^{'}||A_{2}^{'}|=\beta_{ij}|A_{1}||A_{2}|
  \end{eqnarray}
  \begin{eqnarray}\label{15}
   \sum_{i=1}^{2}(|A_{i}^{'}|^{2}-\alpha_{ii}|A_{i}|^{2})U_{ii}^{'}=0
 \end{eqnarray}
 Where $ \beta_{12}=\gamma\dfrac{U_{12}-v}{U_{12}^{'}} $. It is easy to show that $ \beta_{12}=\alpha_{12} $ and then the equations (\ref{12}), (\ref{14})  are equivalent.The equations (\ref{13}) and (\ref{15}) are linear equations for two variables $ |A_{1}^{'}|^2 $ and $ |A_{2}^{'}|^2 $ that their solutions are as follows:
\begin{eqnarray}\label{16}
|A_{i}^{'}|^{2}=\alpha_{ii}|A_{i}|^{2}~~~;~~~i=1,2
\end{eqnarray}
 But the above solutions are inconsistent with equation (\ref{12}), because the direct substitution of equations (\ref{16}) into equation (\ref{12}) leads to the following incorrect equality (see figure 1):
 \begin{eqnarray}\label{17}
 \dfrac{\alpha_{11}\alpha_{22}}{\alpha_{12}^{2}}=1
 \end{eqnarray}
So the system of equations (\ref{12}) and (\ref{13}) are not consistent with equations (\ref{14}) and (\ref{15}). In the other word there is not transformation for the  wave function that leads to correct transformation for Born probability density. Therefore the Born interpretation is incompatible with the Lorentz symmetry of Salpeter equation.
 \begin{figure}
 \includegraphics[scale=0.8]{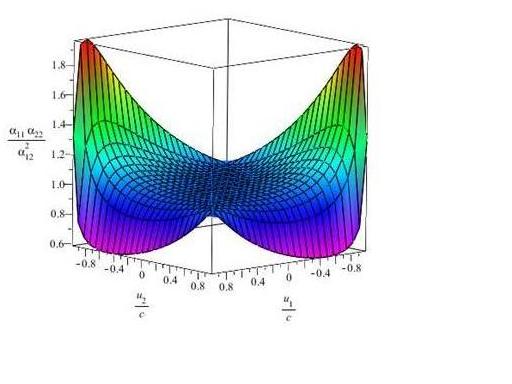}
 \caption{ The diagram $\dfrac{\alpha_{11}\alpha_{22}}{\alpha_{12}^{2}}  $ is plotted versus $ u_1 $ and   $ u_2 $ ( which $ u_1 $ and   $ u_2 $ are the corresponding speeds  of the momentums $ p_1 $ and   $ p_2 $ respectively . It is clear that ($\dfrac{\alpha_{11}\alpha_{22}}{\alpha_{12}^{2}} $ ) is not equal to unity. So the Born rule is inconsistent with the Lorentz symmetry of the Salpeter equation.  It should be noted that if $\dfrac{|u_{1}-u_{2}|}{c}\ll 1$ then  $\dfrac{\alpha_{11}\alpha_{22}}{\alpha_{12}^{2}}\approx1  $, even if the speeds  $u_{1}$ and $u_{2}$ are comparable
 to the speed of ligh.
  }
 \label{fig:1}
 \end{figure}
  \section{ The Correction of the Born Rule}
In the previous section we show that the Born rule is incompatible with the Lorentz symmetry of Salpeter equation, so to make a consistent formalism for relativistic quantum mechanics, the Born rule or Salpeter equation (or both of them) should be modified. In fact by using the Dirac equation, the Born rule is preserved and the Salpeter equation is modified. Another possibility is to keep the Salpeter equation and modify the Born rule. In this regard as an initial effort, the following general form for the relativistic correction for the Born rule is recommended (in one diminution):
\begin{eqnarray}\label{18}
\rho=|\hat{A}\psi|^{2}+|\hat{B}\psi|^{2}
\end{eqnarray}
In which the pseudo-differential operators $\hat{A}$ and $\hat{B}$  are define as follows:
\begin{eqnarray}\label{19}
\hat{A}  =\sqrt{\hat{E}+1}
\end{eqnarray}
\begin{eqnarray}\label{20}
\hat{B}=\frac{\hat{p}}{\sqrt{\hat{E}+1}}
\end{eqnarray}
For understanding the origin of the above definitions, consider the general positive energy solution of Dirac equation (in one dimension:
\begin{eqnarray}
\Psi(x,t)=\begin{bmatrix}
\psi_{1}\\
\psi_{2}
\end{bmatrix}=\frac{1}{\sqrt{2\pi \hbar}}\int_{-\infty}^{+\infty}\varphi(p)u(p) e^{\frac{i}{\hbar}(px-E(p)t)}dp
\end{eqnarray}
where $ \Psi(x,t) $  represents the two-component Dirac spinor and $\varphi(p)$ is wave function in momentum space and $u(p)$ is positive energy plane wave solution of Dirac equation:
\begin{eqnarray}\label{22}
u(p)=\begin{bmatrix}
\sqrt{E+1}\\
\frac{p}{\sqrt{E+1}}
\end{bmatrix}
\end{eqnarray}
If we define $\psi$ as:
\begin{eqnarray}\label{23}
\psi(x,t)=\frac{1}{\sqrt{2\pi \hbar}}\int_{-\infty}^{+\infty}\varphi(p) e^{\frac{i}{\hbar}(px-E(p)t)}dp
\end{eqnarray}
Then the Dirac spinor,$\Psi$, could be rewritten in terms of  $\psi$as follow:
 \begin{eqnarray}\label{24}
\Psi(x,t)=\begin{bmatrix}
\hat{A}\psi\\
\hat{B}\psi
\end{bmatrix}
\end{eqnarray}
Also the Dirac probability density, $\rho_{D}$, and current density, $\textbf{J}_D$,could be rewritten in terms of  $\psi$ as follow:
\begin{eqnarray}\label{25}
\rho_{D}=(\hat{A}\psi)(\hat{A}\psi)^{\star}+(\hat{B}\psi)(\hat{B}\psi)^{\star}
\end{eqnarray}
\begin{eqnarray}\label{26}
J_{D}=(\hat{A}\psi)^{\star}(\hat{B}\psi)+(A\psi)(\hat{B}\psi)^{\star}
\end{eqnarray}
From eq.(\ref{23}) it is clear that $ \psi(x,t)$ satisfies the Salpeter equation, Therefore the Salpeter equation  along with the interpretation of the wave function based on the eq. (\ref{18}) , leads to a formalism equivalent with the positive energy solutions of the Dirac equation. of course we know that $\rho_D $ and $\textbf{J}_D $ make altogether a four-vector; so the eq. (\ref{25}) and eq.(\ref{26}) can be used as an acceptable relativistic interpretation of the Salpeter equation.
 It shows the possibility of a consistent probabilistic interpretation of the Salpeter equation. Our purpose in this paper is only to show such possibilities; although other possibilities for a correct relativistic interpretation of the wave function might exist. However to achieve a consistent relativistic quantum mechanics based on the Salpeter, the probability density of position must be necessarily deviated from the Born rule.

The question that arises at this stage is: The deviation from the Born rule in the relativistic level is just only for the probability distribution of position or the probability distribution of the other quantities deviated from the Born rule? To answer this question, we note that there are several methods for extracting the Born rule for other quantities from this rule is in the case of position. As a specific example,$ |\varphi|^2 $as the momentum probability density have been obtained by analyzing the time of flight measurement with the assumption of the Born rule in position measurement\cite{26}. In general, the establishment of Born rule on other observable quantities can be derived according to the causal theory of measurement with the assumption Born rule in position measurement \cite{26,27}. In all such demonstration, the measurement of other quantities related to measurement of position and thus the establishment of the Born rule for the probability density of position has a key role in the derivation of this rule for other observable. Consequently, deviation of position probability density from the Born rule can deviate probability density of other quantities from the Born rule.

\section{CONCLUSION and Discussion}
We show that the Square Root Klein-Gordon equation with a reformation of the Born rule will be Lorentz invariant.Our calculations in this paper were performed in the absence of external field, but the most doubts about the Lorentz invariance of Salpeter equation are in the presence of external fields \cite{28,29,30,31}.For example in 1963,J.Sucher  showed that the Salpeter equations not Lorentz invariance in the presence of interactions by entering the interaction with minimal coupling $ \partial_{\mu}\longrightarrow \partial_{\mu}-ieA_{\mu} $. Sucher assumed that the wave function is a scalar\cite{29}. But the default scalar wave function is not required and the transformation properties of the wave function may be more complex. In fact Lorentz  transformation of the wave function must be determined according to its physical interpretation and we have shown that with a proper interpretation of the wave function the Salpeter(in the absence of the external field) will be Lorentz invariant off course in this case, wave function is not scalar. So in the presence of interaction, it is possible the wave function may not be scalar. Therefore the Sucher proof should not be considered as a final proof of the non-Lorentz invariance of the square root Klein- Gordon equation and may be, this equation is Lorentz invariant with the proper interpretation of the wave function even in the presence of interaction. We leave the checking of this possibility as a open problem.

\end{document}